# Magnetic field suppression of Andreev conductance at superconductor–graphene interfaces

Piranavan Kumaravadivel*[1,‡], Scott mills*[1], and Xu Du[1†]

[1]Department of Physics and Astronomy, Stony Brook University, New York 11794-3800, USA.

Studying the interplay between superconductivity and quantum magnetotransport in two-dimensional materials has been a topic of interest in recent years. Towards such a goal it is important to understand the impact of magnetic field on the charge transport at the superconductor-normal channel (SN) interface. Here we carried out a comprehensive study of Andreev conductance under weak magnetic fields using diffusive superconductor- graphene Josephson weak links. We observe that the Andreev conductance is suppressed even in magnetic fields far below the upper critical field of the superconductor. The suppression of Andreev conductance depends on and can be minimized by controlling the ramping of the magnetic field. We identify that the key factor behind this suppression is the reduction of the superconducting gap due to the piling of vortices on the superconducting contacts. In devices where superconducting gap at the superconductor-graphene interface is heavily reduced by proximity effect, the enlarged vortex cores overlap quickly with increasing magnetic field, resulting in a rapid decrease of the interfacial gap. However, in weak links with relatively large effective superconducting gap the AR conductance persists up to the upper critical field. Our results provide guidance to the study of quantum material-superconductor systems in presence of magnetic field, where 'survival' of induced superconductivity is critical.

[*] These authors contributed equally to this work.

[‡] Present address: School of Physics and Astronomy, University of Manchester, Manchester M13 9PL, UK.
[†] Email: xu.du@stonybrook.edu

Superconducting weak links on two-dimensional electron systems (2DES) have been extensively studied for exploring many of the emergent phenomena in condensed matter physics. In recent years there has been a growing interest in studying these structures in the presence of relatively strong magnetic fields for understanding the interplay between quantum Hall edge states and superconducting correlations. With the advent of graphene and a plethora of other 2D materials and topological insulators, the combination of chiral edge states and superconductivity holds promise in the study of novel phenomena such as Majorana fermions, non-abelian anyons, quantum Hall edge state supercurrent, and Andreev conversion of QH edge states[1-8]. Experimentally, the delicate nature of these phenomena requires devices of the highest quality in both the 2DES channels and superconductor-normal metal (SN) interface, as well as superconductor electrodes that can retain superconducting correlations in high magnetic fields. However, due to the emergence of Meissner and vortex phases in superconducting thin films and type II superconductors in the presence of magnetic fields, the charge transport at sample specific SN interface can be significantly complicated. Such effects have rarely been discussed in previous works. A careful systematic study of intrinsic and extrinsic factors affecting charge transport in these devices is therefore of significant importance. Charge transport in SN junctions takes place primarily via the Andreev reflection (AR) process: an electron (hole) enters the superconductor from the normal side and gets retro-reflected as a quasiparticle hole (electron) so that a Cooper pair can form inside the superconductor. The coherent propagation of these phase conjugated quasiparticles enhances the conductivity of the SN interface. In the diffusive limit and in low magnetic fields prior to the formation of Landau levels and cyclotron orbits, because the incident/reflected quasiparticles follow the same trajectory on the normal side of the interface they are immune to phase breaking effect by magnetic fields. On the other hand, the impact of the magnetic field on Andreev reflection may be expected [9, 10] considering the presence of screening currents on the surface of the superconductor in magnetic field. Screening currents are composed of a moving Cooper pair condensate and in order to accommodate this Cooper pair momentum, the incident and Andreev-reflected quasiparticles also acquire a momentum shift at the SN interface. When the applied magnetic field is sufficient that the associated energy shift is comparable to the superconducting gap at the SN interface, Andreev reflection probability becomes significantly suppressed, diminishing the conductance enhancement. Besides the above "Doppler shift" scenario, the presence of superconducting vortices may also play an important role in the charge transport at the SN interface and has not previously been studied.

Experimental study of AR in magnetic field has been previously carried out with niobium-semiconductor 2DEG junctions [11] . There it was observed that at low magnetic field of a few 100 mT (well below the upper critical field of niobium $B_{c2}$~2 T), the AR is almost completely suppressed. The suppression was explained using the Doppler shift model, considering only the diamagnetic Meissner currents in the superconducting leads which rapidly suppresses the zero-bias AR conductance with increasing magnetic field. Despite the qualitative agreement, the Doppler shift model has several major discrepancies with the experimental observations. First of all, the model predicts that the screening currents broaden the energy (bias voltage) range of the gap features in the differential conductance[10, 12] that has not been observed in experiments. Secondly, it is established that in superconductor thin films, magnetic flux lines begin penetrating and forming vortices in extremely low fields [13]. Both the distribution of the screening current and local order parameter should therefore be affected by how these vortices are distributed on superconducting films. The distribution will in turn depend on the dynamics by which the vortices enter and exit the superconducting thin films when ramping the magnetic field up or down to the desired value. The impact of the superconducting vortices on Andreev reflection therefore should be considered when exploring magneto-transport in SN junctions. Moreover, as shown by some recent works, including our own observations discussed below, the magnetic suppression of the AR appears to be sample dependent[11, 14]. This indicates that such phenomenon may not be intrinsic but instead strongly affected by certain characteristics of the individual devices. Unraveling these effects and exploring ways to preserve the superconducting coherence can therefore be a useful guide for future investigations on the interplay between superconductivity and quantum magnetotransport phenomena.

In this work, we carried out a comprehensive study on charge transport in superconductor-graphene-superconductor (SGS) Josephson weak links in the presence of weak magnetic fields. A suppression of AR conductance is observed even in magnetic fields far below the upper critical field ($B<<B_{c2}$). The dependence of the AR conductance on the ramping dynamics of the magnetic field reveals the important role of vortices and vortex pinning. The key factor behind the rapid AR suppression is identified to be the strongly reduced superconducting gap at the superconductor-graphene (SG) interfaces compared to that of the bulk superconducting leads. As a result the superconducting coherence length, and hence the size of the vortex cores are enlarged. Combined with vortex pinning, the overlapping of the vortices rapidly reduces the effective superconducting gap and thereby the AR conductance. By improving the SG interface, we can optimize the effective superconducting gap to reach a value closer to the intrinsic BCS gap of the superconducting leads. In these devices the impact of the vortices on AR is minimized and the AR conductance persists closer to the upper critical field of the superconducting contacts.

**Methods**

SGS Josephson weak links are fabricated on $SiO_2$/Si substrates using mechanically exfoliated highly oriented pyrolytic graphite (HOPG). A buffer layer is deposited between graphene and the superconductor to facilitate both good adhesion and improve charge transmission between graphene and the superconducting contacts. Four types of buffer layers: Ti (1 nm), Ti (2 nm)/Pd (1.5 nm), Ti (2 nm)/Au(2 nm), and V(2 nm) are tested, by thermal (e-beam) evaporation in a UHV environment. Immediately after the evaporation and without breaking vacuum, superconducting thin film of Nb or NbN is coated onto the samples via DC magnetron sputtering. For Nb thin films, sputtering is done in pure Ar environment[15]. For NbN thin films, reactive DC Magnetron sputtering is carried out in a mixture of $N_2$ and Ar [16]. All samples have graphene channels that are of length ~ 0.7 μm and width ~1.5-10 μm in width (Figure 1A). The mobility of the graphene channel is estimated to be $\mu = \sigma/ne$ ~5000-6000 cm$^2$/Vs from two-terminal conductivity just below the transition temperature ($T_C$) of the superconducting leads ($T_C$ ~ 11 K for NbN and ~8.5 K for Nb). The mean free path is calculated to be $l_{mfp} = \frac{\sigma \hbar}{e^2}\sqrt{\frac{\pi}{n}}$ ~ 50-60 nm (see Supplementary Information).

**Results**

Basic characterizations of the samples are presented in Figure 1B. At low temperatures $T<<T_C$, all devices show supercurrent or a precursor of supercurrent through a vanishing or sharply reduced differential resistance (*dV/dI*) at zero bias current. The strong Josephson coupling indicates a highly transparent interface between graphene and the superconducting contacts. Besides supercurrent, clear evidence of multiple Andreev reflections (MARs) is observed from the

*dV/dI* versus bias voltage(*V*<sub>bias</sub>) curves, as shown from an NbN-graphene device at $T = 0.4$K in Figure 1B. The valleys of the differential resistance oscillations appear at the expected values for MARs at $V_{bias} = \frac{2\Delta_{eff}}{Ne}$, where $\Delta_{eff}$ is an *effective* energy gap and *N*=1,2,3.. is an integer. We notice that $\Delta_{eff}$ is sample dependent and is usually significantly reduced from the bulk superconducting gap of the leads (~1.3 meV for Nb and ~1.7 meV for NbN, estimated using BCS theory $\Delta_{BCS} = 1.76 k_B T_C$ with *Tc* ~8.5 K for Nb and 11 K for NbN as measured in our samples. For example for the NbN junction shown in Figure 1B, $\Delta_{eff}$~0.15 meV; while in all our other devices, $\Delta_{eff}$ ranging between 0.3~0.95 meV is observed. In our analysis, we used the outermost valley in the *dV/dI* vs. bias curves to identify the value of *2Δ*<sub>eff</sub>. This is based on the theoretical calculations of the MAR spectrums in both ballistic and diffusive SNS weak links[17-19], where the outermost *dV/dI* valley (i.e., conductance peak) appear to give a good estimation to the value of the superconducting gap. In addition, while temperature and disorder may affect the accuracy of the gap values, because all the measurements were carried out at the same temperature and in devices with similar mobility, the parallel comparisons between different samples and in different magnetic fields (discussed later) are still reasonable. With these values of $\Delta_{eff}$, the coherence length in our diffusive devices is estimated to be $\xi = \sqrt{\frac{\hbar D}{\Delta_{eff}}} = \sqrt{\frac{\hbar v_F l_{mfp}}{\Delta_{eff}}} \sim 200-500$ nm which is slightly less than the junction length.

The reduction of superconducting gap is commonly observed and reported in SNS weak links[15, 20-22]. The reduced superconductor pairing potential can be a result of either proximity effect at the SN interface in presence of the buffer layers[23], or interfacial mixing/diffusion between the superconductor and the buffer layer. In any case we found that such gap reduction can be minimized by reducing the thickness of the buffer layer. The largest effective gap was achieved in Nb-Ti(1 nm)-G samples. In these samples, Ti does not form a continuous thin film but instead islands. The Ti islands aid in the mechanical adhesion of the superconducting contacts and charge transmission takes place predominantly between the superconductors and graphene where Ti is absent. As a result, a larger effective gap of $\Delta_{eff}= \sim 0.6 - 0.95$ meV is routinely observed. The remaining gap reduction is presumably due to the antiproximity effect from the presence of graphene at the SG interface. The effective gap of ~0.95 meV is comparable to the bulk gap of Nb, and is consistent with the highest values reported in similar SGS Josephson weak links[6, 8].

Next we focus on the characteristics of AR in presence of weak magnetic field (i.e., $B << \frac{1}{\mu} \sim$ 1-2 T). The cyclotron orbit and Landau levels are not formed in the diffusive graphene samples and magnetotransport is classical, enabling us to focus on the impact of magnetic field on the SN interface. The main results are summarized in Figure 2 with *dV/dI* values normalized by the normal resistance of the junction ($R_N$) just below *Tc*. When a magnetic field is applied after the samples are zero-field cool (ZFC)-ed below $T_C$, the AR-associated gap feature in the differential resistance curve becomes suppressed. In some of the samples (e.g., the NbN-Pd/Ti-G sample shown in Figure 2A) the oscillatory MARs features become completely suppressed under a very small magnetic field, less than 10 mT. Further, monotonously increasing the magnetic field to different values at a fixed ramp rate and measuring the *dV/dI* as a function of $V_{bias}$ we find that the AR enhancement of conductance (~20% at $V_{bias}$ ~2$\Delta_{eff}$ at B=0) quickly reduces and eventually vanishes around B=200mT which is much lower than the upper critical field of NbN ($B_{C2}$ > 10 T). The magnetic suppression of AR conductance appears to be sample dependent. For example in the Nb-Au/Ti-G sample shown in Figures 2A, the AR conductance enhancement remains observable in magnetic field B~1 T. In particular with the Nb-Ti(1 nm)-G sample, the AR conductance is only very weakly affected by the magnetic field and persists close to the upper critical field ($B_{C2}$ ~ 2 T at T = 4.2K). The effect of magnetic field on single ARs at the SN interface can be better evaluated using the excess current ($I_{exc}$). The excess current is obtained by extrapolating the normal section of the *IV* curve and identifying its intersection on the current axis at zero bias. The excess current contains information on Andreev reflection and is insensitive to decoherence compared to supercurrent. $I_{exc}R_N$ at various ZFC-ed field values is shown in Figure 2B. Evidently the magnetic suppression of AR varies significantly in different samples.

Along with the magnetic suppression of the AR conductance, a suppression of the effective superconducting gap is also observed in all our samples. This is evident from the width of the sub gap valley feature in the *(1/$R_N$) dV/dI*_ versus $V_{bias}$ plots shown in Figure 2A. For samples with large effective gaps, we can reliably obtain the values of the effective gap from the sharp kink in the *dV/dI* curve at $V_{bias} = 2\Delta_{eff}$ . As shown in Figure 2A for a Nb-Ti(1nm)-G sample, the effective gap decreases with increasing magnetic field to 100 mT then to 1 T. For samples with small effective gap, it is difficult to extract the effective gap in magnetic field because of the rather featureless "V"-shaped *dV/dI* curves. Nevertheless, one can clearly see that the width of the "V"-shaped valley decreases with increasing magnetic field.

To identify the origin of the strong magnetic suppression of AR, several possible factors are considered. First AR is affected by the charge transmission properties of the SG interface which may depend on magnetic field. The various buffer layers studied here give rise to different transparencies. We found that both Ti/Pd and Ti/Au buffer layers offer excellent and reliable charge transmission with graphene, indicated by the strong zero-field and zero-bias conductance enhancement. A very thin (discontinuous) layer of Ti gives reasonable transparency, although less transparent compared to that in the Ti/Pd and Ti/Au buffered samples. The V buffer layer generally yields large stress and poor interface transparency. But overall the samples are still weak-link-like (as opposed to be "tunneling"-like where supercurrent is absent and the resistance shows a maximum when $V_{bias}$ is within the superconducting gap and quasiparticle tunneling is suppressed). Despite the vast qualitative differences in their charge transmission, a comparative study of all the buffer layers shows no systematic dependence of the suppression rate on the interface transparency.

Secondly, the bulk superconductor gap (1.3 meV for Nb and 1.7 meV for NbN) of the contacts also does not show a systematic influence on the suppression rate of AR in magnetic field. However, for devices with relatively large *effective* superconducting gap (i.e., a broader AR gap feature in the *dV/dI vs.* $V_{bias}$ curve), AR is consistently less susceptible to the magnetic field. As shown in Figure 2B, the rate of the magnetic suppression of AR with increasing magnetic field has a clear monotonic dependence on the width (in $V_{bias}$) of the sub-gap conductance in the *(1/ $R_N$) dV/dI vs.* $V_{bias}$ curves for the different samples studied. At fixed low magnetic field, temperature appears to play little role on the magnetic field suppression of AR when it is well below $T_C$. Figure 2C shows a comparison between the magnetic field dependence of $I_{exc}R_N$ measured at 0.4 K and 4.2 K, for devices with Ti/Pd and V buffer layers and Nb contacts ($Tc \sim 8.5$ K). In both cases, the excess current follows qualitatively the same dependence on magnetic field, practically independent of the temperature. Far below $T_C$, such weak temperature dependence of $I_{exc}R_N$ is in qualitatively agreement with the BTK model (see Supplementary Information).

Thirdly, we explored the impact of the dynamics of the magnetic field on AR conductance. Besides ZFC, we studied two other sequences: one is the field-cool (FC) process, where a sample is cooled down below $T_C$ *after* a magnetic field is applied and the other is the "down-ramping" (DR) procedure, magnetic field is ramped up from zero at $T$ ($<< T_C$), first to a high value (B > 1 T) and then decreased back down to the desired value where $dV/dI$ as a function of $V_{bias}$ is measured. As for the ZFC, a fixed ramping rate is maintained for all measurements. Figure 3 shows a comparison of the AR related features the NbN- and Nb-based devices under the different magnetic field ramping sequences. For a given low field between 10~700 mT, NbN-based samples show significantly larger dip in the sub-gap differential resistance and hence a higher $I_{exc}R_N$ for both ZFC and DR procedures, compared to that in the ZFC procedure. In particular, the DR procedure allows the AR enhancement of conductance to persist up to ~1 T. For Nb-based devices the DR procedure similarly allows AR to be less susceptible to magnetic field for $B$ < 200 mT. However the difference between ZFC and DR is less significant compared to that for the NbN-based devices. We note that magnetic hysteresis from the superconducting magnet has negligible role in these observations.

**DISCUSSION**

The observation of the dynamics-dependent magnetic suppression of AR suggests the important role of superconducting vortices in these measurements. Indeed it is established that for thin film superconductors, vortices form a stable state once the magnetic field is above a critical value of the order of $B_m \sim \frac{\Phi_0}{L^2}$, where $\Phi_0 = h/2e$ is the magnetic flux quantum and $L$ is the width of the superconducting thin film[13]. For the geometry of our devices ($L$~1-2 µm), $B_m \sim 1$ mT, which is at the very low end of the magnetic fields applied here. Furthermore, the dynamics of the vortices is different when entering and exiting the superconducting pads and it directly affects their spatial distribution[24, 25]. With increasing magnetic field and in the case of ZFC, vortices tend to pile up at the edge of the superconducting thin films (where they enter the thin film) due to pinning. On the other hand, when the magnetic field decreases (in the case of DR), the vortices at the edge of the superconducting film rapidly exit from the superconductor, leaving a much lower density regime for vortices at the edge. In the case of FC, the vortices are formed during the superconducting transition and distribute more uniformly inside the superconductor. The difference in the vortex density distributions between ZFC and FC, as well as between ZFC and DR is expected to be stronger for a strong-pinning superconductor (such as NbN) than for a relatively weak pinning superconductor (such as Nb). This is consistent with our observations where much stronger hysteresis in the $I_{exc}R_N$ vs. magnetic field was observed for the NbN-based device (Figure 3B) compared to that for the Nb-based devices (Figure 3D).

The effect of magnetic field on AR in relation to the spatial distribution of the vortices can be explained considering the strong current crowding effect[26] at the SG contacts as illustrated in Figure 4D. In this simplified picture where the SG interface is modeled as a transmission line of resistor network with uniformly distributed contact resistance and sheet resistance, the current $I$ flows from the superconductor to graphene with spatial distribution $j(x) = \frac{a}{L} I \frac{\cosh(ax/L)}{\sinh(a)}$ where $a = \sqrt{\frac{R_S}{R_C}}$, $L$ is the width of the superconducting contact, $R_C$ is the interfacial resistance between graphene and superconductor, and $R_S$ is the resistance of graphene underneath the contact. In a high transparency contact: $R_c \ll R_s$, current flows from the contacts into graphene primarily at the inner edge of the contacts ($x \approx L$). It is in this region vortices enter and exit the superconducting contacts, with their density determined by the magnetic field ramping procedure. During ZFC-ed field measurements the vortices are denser at the edges. When cycled back to the same field by the DR procedure, the edges have a lower vortex density compared to the ZFC for the same field. As a result, one expects a strong magnetic field and ramping dynamics dependence in charge transport characteristics.

With both vortex and current crowding at the inner edges of the superconducting contacts, the suppression of AR can be explained by the magnetic field dependence of the averaged superconducting gap at the SG interface. For each vortex, the superconducting order parameter decreases towards the vortex core over a distance of $\xi$, the superconducting coherence length. With increasing magnetic field and hence increased vortex density at the edges, the vortices become increasingly overlapped and the average order parameter in the superconductor decreases. This is reflected in the decreasing effective gap with increasing magnetic field as observed in our *dV/dI* *(V_{bias})* measurements (Figure 4A). The gap reduction is expected to be more sensitive to magnetic field when the zero-field gap is small and therefore $\xi = \sqrt{\dfrac{\hbar D}{\Delta_{eff}}}$ (or the vortex core size) is large. As a result the vortices overlap and reduce the average superconducting gap more quickly compared to when the effective superconducting gap is larger.

To highlight the impact of the effective superconducting gap on AR we plot $I_{exc}R_N$ versus the effective gap in various magnetic fields in Figure 4B, taken from the Nb-Ti-G sample. While both $I_{exc}R_N$ and the effective gap show hysteretic magnetic field dependence, the relation between the excess current and the effective gap is non-hysteretic and linear within the experimental uncertainty. The linear dependence which extrapolates to the origin of the plot is qualitatively consistent with the theories[17, 27, 28] on the gap dependence of the $I_{exc}R_N$ (see Supplementary Information). Besides excess current, we also compare the bias dependence of the differential resistance under ZFC and ZFC-down ramps. It is found that the line shape of differential resistance taken at different ramping procedures closely match with each other whenever they have the same effective gap. Our observations suggest that for a given device, the magnetic suppression of AR conductance is predominantly caused by the suppression of the effective gap by the magnetic field.

While our result suggests that the vortex suppression of gap energy plays a critical role in the AR suppression, it does not rule against the contribution from the "Doppler shift" model [9] especially in the very low magnetic field regime where vortex density is very low. The Doppler shift model considers AR process between normal electrons and diamagnetic supercurrent which leads to a shift in the canonical momentum $e\vec{A} = \mu_0 e \lambda^2 \vec{j}$ . Here $\lambda$ is the London penetration depth and $\mu_0$ is the magnetic constant. As a result the superconducting gap "seen" by the normal charges at the Fermi level is shifted by $\tilde{\varepsilon} \sim e v_F \mu_0 \lambda^2 j$ . This energy shift effectively reduces the zero-bias AR conductance. The magnitude of such Doppler effect is determined by the diamagnetic current density at the SN interface, which increases linearly with magnetic field before the inclusion of vortices. On the other hand, in presence of high vortex density the net diamagnetic current density is largely determined by the vortex density gradient and does not increase beyond the critical current. Considering proximity effect at the SN interface, the local critical current density is expected to be significantly reduced compared to the bulk superconductor. Hence the low diamagnetic supercurrent density at the SN interface only has minor contribution to the magnetic suppression of the AR.

In summary, magnetic suppression of Andreev conductance is observed in diffusive SGS Josephson weak links in weak magnetic fields. The suppression depends both on the magnitude and the ramping procedure of the magnetic field. We identify the key factor behind the magnetic suppression of Andreev conductance to be the suppression of superconducting gap from the piling of vortices. Due to the proximity reduction of the effective superconducting gap at the SN interface, ξ and hence the vortex cores are enlarged, resulting rapid decrease of interfacial gap with increasing magnetic field. In weak links with relatively large effective superconducting gap the AR persists approaching $B_{c2}$. Our work established an understanding of the charge transport across SN interfaces in presence of magnetic field. Moreover, it provides useful guidance for the fabrication and characterization of quantum material-superconductor systems, where study of the interplay between superconductivity and novel quantum phenomena requires "survival" of superconductivity in presence of magnetic field.

**Acknowledgement**

The author would like to acknowledge Dmitri Averin and Konstantin Likharev for insightful discussions, Laszlo Mihaly for support with cryogenic facilities, and Peter Stephens for providing single crystal HOPG. This work was supported by AFOSR under grant FA9550-14-1-0405.

# FIGURES

FIGURE 1.

**A**  **B**

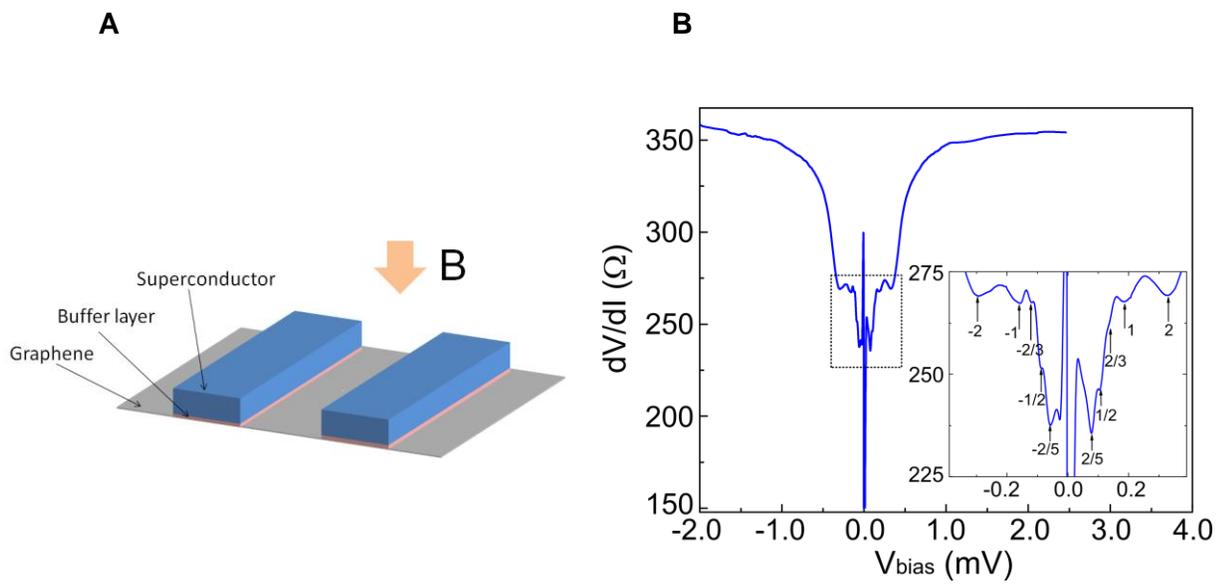

FIGURE 2

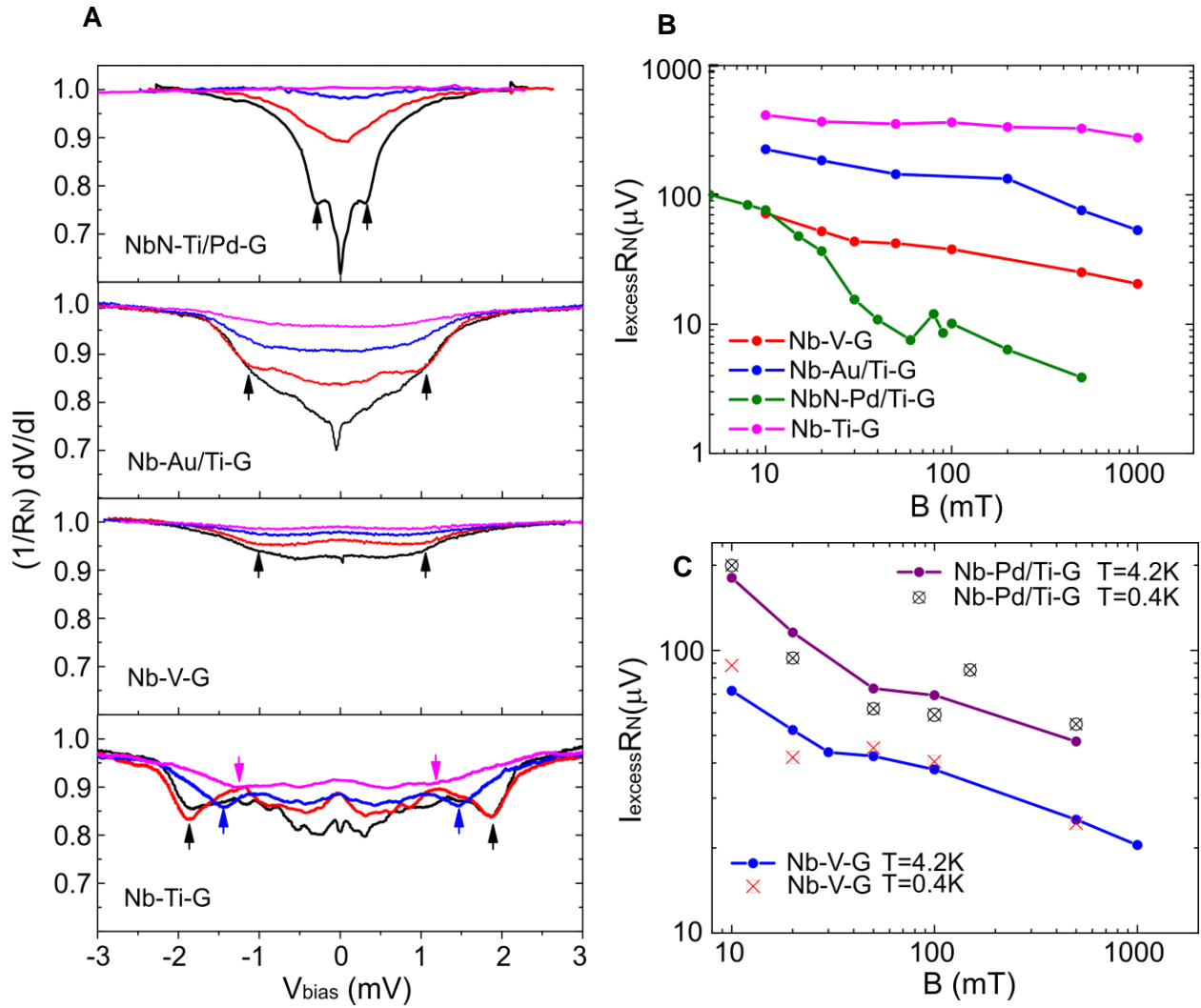

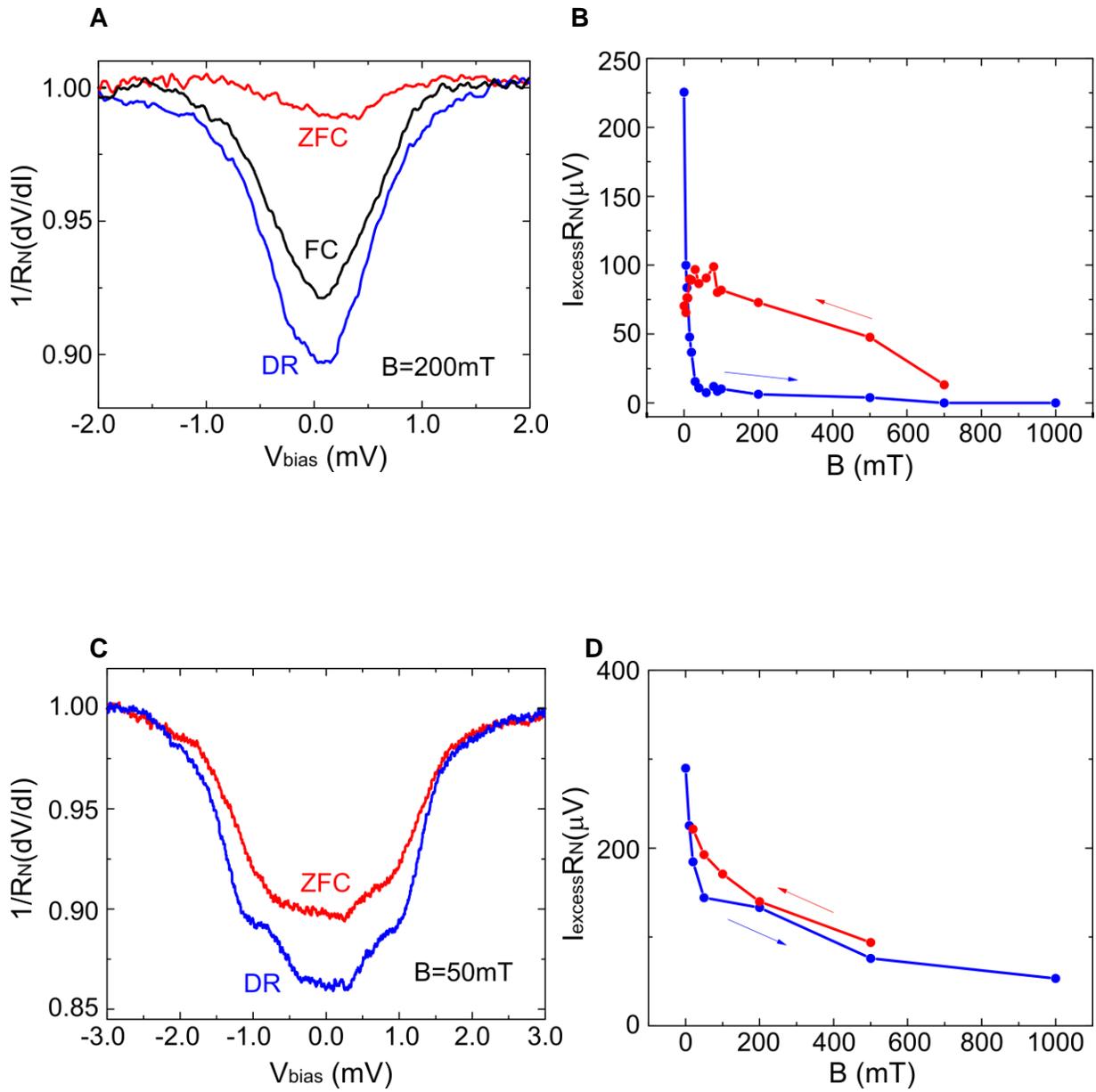

FIGURE 4.

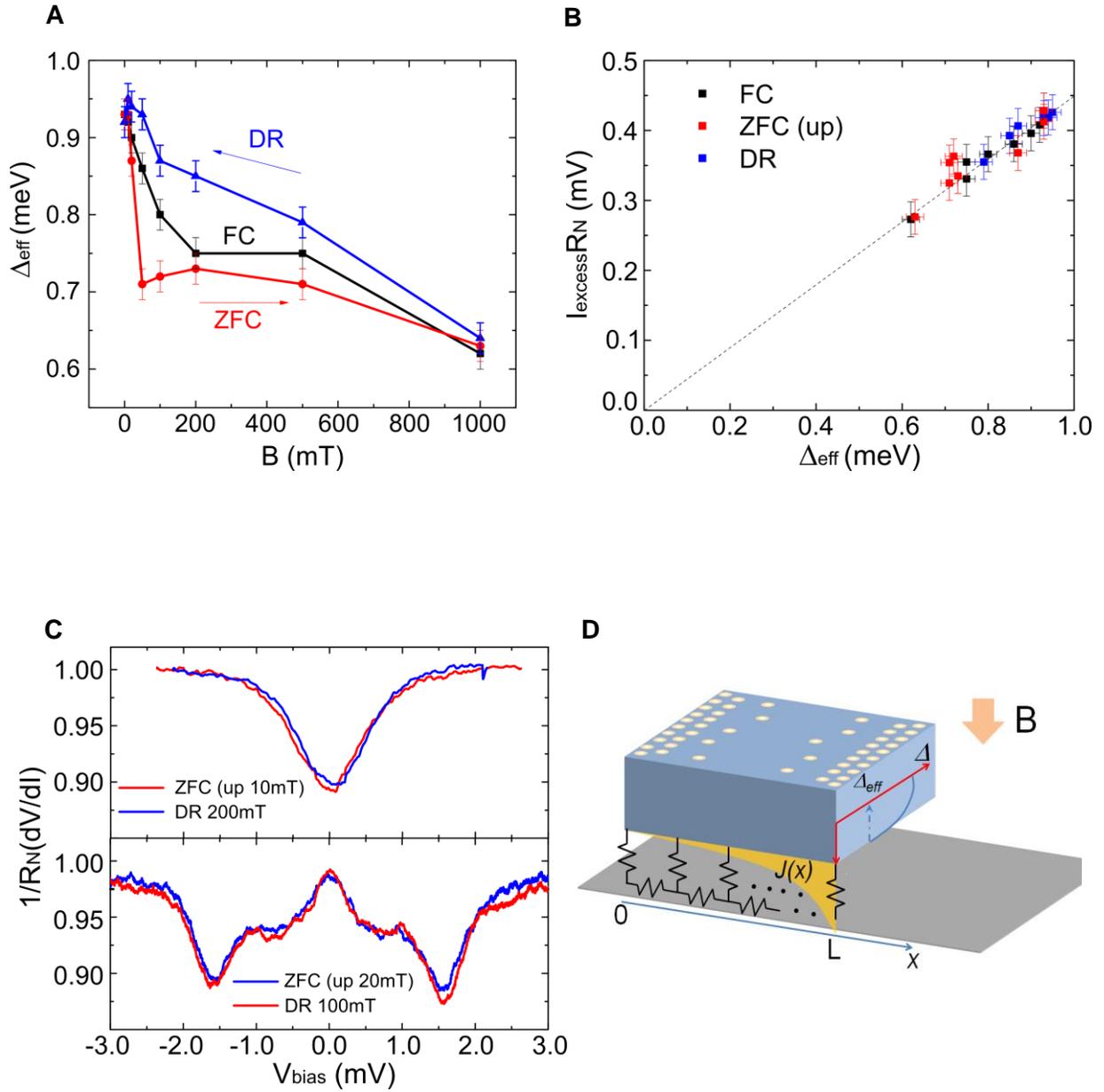

**FIGURE 1: Device Characteristics. (a)** Device geometry and magnetic field direction. **(b)** A typical differential resistance (dV/dI) measurement at $T$ = 0.4K showing subharmonic gap structures (indicated by arrows) at $V_{bias} = \frac{2\Delta_{eff}}{Ne}$ due to MAR and supercurrent. Here $\Delta_{eff}$ is the effective superconducting gap.

**FIGURE 2: Suppression of Andreev Reflection in Magnetic Field. (a)** Dependence of normalized differential resistance $1/R_N(dV/dI)$ versus bias voltage ($V_{bias}$) on applied magnetic field for different buffer layers (from top down): Ti (2 nm)/Pd (1.5 nm), Ti(2 nm)/Au(2 nm), V(2 nm) and Ti (1 nm), measured at $T$ = 4.2K. Magnetic field B =0 (black), 10(red), 100 (blue) and 1000 (pink) mT. The first inflection point after $dV/dI$ begins to decrease, where $V_{bias} = 2\Delta_{eff}$, is denoted by the arrows. **(b)** $I_{exc}R_N$ dependence on applied magnetic field at $T$ = 4.2K for the samples in **(a)**. The samples with a larger effective gap has a weaker dependence on magnetic field and shows significant excess current even at B = 1 T. **(c)** Temperature dependence of $I_{exc}R_N$, showing temperature is not a significant factor for the samples measured.

**FIGURE 3: $I_{exc}R_N$ and Andreev Reflection Dependence on Ramping Direction of Applied Field. (a)** $(1/R_N)$ $dV/dI$ versus $V_{bias}$ for zero field cooled (ZFC) up-ramp, field cooled and zero field cooled down-ramp at $B$ = 200 mT at $T$ = 4.2K for the NbN sample from (Figure 2). **(b)** $I_{exc}R_N$ versus magnetic field for ZFC up and down ramp (DR) for the sample in **(a)**, the larger hysteresis is attributed to the stronger vortex pinning in NbN compared to Nb. **(c)-(d)** Same as **(a)-(b)** for the Ti/Au/Nb sample from (Figure 2), showing the weaker dependence on ramping direction characteristic of the Nb samples compared to NbN.

**FIGURE 4: Effective Gap Dependence on Applied Field. (a)** Dependence of the effective gap on the ramping sequence of the applied field for the Nb-Ti-G sample. **(b)** Linear relationship between $I_{exc}R_N$ and effective gap. **(c)** Matching of $(1/R_N)$ $dV/dI$ curves with the same value of excess current, but different procedures: ZFC and DR, for applying the magnetic field. **(d)** Transmission line model of superconducting contacts with vortices piling along the edge of the interface. The current from the lead to the sample is concentrated near the edge, so the effective gap seen during AR changes when vortices enter the lead.